 \documentclass[final,3p,times]{elsarticle}
\usepackage{latexsym}
\usepackage{amssymb,latexsym} 
\usepackage[fleqn]{amsmath}
\usepackage{graphicx}% Include figure files
\usepackage{dcolumn}% Align table columns on decimal point
\usepackage{bm}% bold math
\usepackage{caption}
\usepackage{subcaption}
\usepackage{grffile}
\usepackage{multirow}
\usepackage[colorlinks=true,allcolors=blue]{hyperref}
\usepackage[mathlines]{lineno}% Enable numbering of text and display math
\usepackage{amsthm}

\journal{Commun Nonlinear Sci Numer Simulat}
\begin{document}

\begin{frontmatter}



\title{Fractal Measures and Nonlinear Dynamics of Overcontact Binaries}

 \author[label1]{Sandip V. George}
\author[label2]{R. Misra}
 \author[label1,label3]{G. Ambika}
 \address[label1]{Indian Institute of Science Education and Research(IISER) Pune, Pune, India - 411008 }
 \address[label2]{Inter-University Centre for Astronomy and Astrophysics (IUCAA) Pune, Pune, India - 411007}

\address[label3]{Indian Institute of Science Education and Research (IISER) Tirupati, Tirupati, India - 517507}
\ead{g.ambika@iisertirupati.ac.in}

%%\author{}

%\address{}

\begin{abstract}
Overcontact binary stars are systems of two stars where the component stars are in contact with each other. This implies that they share a common envelope of gas. In this work we seek signatures of nonlinearity and chaos in these stars by using time series analysis techniques. We use three main techniques, namely the correlation dimension, $f(\alpha)$ spectrum and the bicoherence. The former two are calculated from the reconstructed dynamics, while the latter is calculated from the Fourier transforms of the time series of intensity variations(light curves) of these stars. Our dataset consists of data from 463 overcontact binary stars in the Kepler field of view \cite{prvsa2011kepler}. Our analysis indicates nonlinearity and signatures of chaos in almost all the light curves.  We also explore whether the underlying nonlinear properties of the stars are related to  their physical properties like fill-out-factor, a measure of the extend of contact between the components of an overcontact binary system .  We observe that significant correlations exist between the fill out factor and the nonlinear quantifiers. This correlation is more pronounced in specific subcategories constructed based on the mass ratios and effective temperatures of the binaries. The correlations observed can be indicative of variations in the nonlinear properties of the star as it ages. We believe that this study relating nonlinear and astrophysical properties of binary stars is the first of its kind and is an important starting point for such studies in other astrophysical objects displaying nonlinear dynamical behaviour.
\end{abstract}


\end{frontmatter}


\section{Introduction}           %% first-level sections will be auto-capitalized
\label{sect:intro}

Our understanding of astrophysical objects has been considerably aided by the use of the dynamical systems theory. Apart from its rich history in uncovering the areas of celestial and planetary dynamics, it has also been put to considerable use in the study of accretion disc physics, tidal capture of binaries, planet dynamics, galaxy simulation models etc \cite{jacob2018recurrence,marling1995roleofchaosBoundary,satyal2013application,manos2014chaos,zotos2012order}. The light variations of many variable stars can be understood better when described using concepts of nonlinear dynamics\cite{reg06}. Among them, pulsating variable stars are the most well studied, where multiple nonlinear dynamical models exist that help to understand the dynamics.  But in general,  most often we rely on the observational data of their intensity variations to understand their dynamics using the tools of nonlinear time series analysis \cite{ste86,buchler1995chaotic,kiss2002period,ambika2003chaotic,plachy2013low}. These tools need long, high quality datasets to arrive at conclusions. With the advent of the Kepler space  telescope, which measures light intensities with high precision and over long periods in time, these needs have been met to some extend. Kepler light curves continue to be used successfully in unravelling nonlinear phenomena, indicating the presence of complex deterministic dynamics in many pulsating variables \cite{szabo2010does,lindner2015strange,plachy2018chaotic}.

While the tools of nonlinear time series analysis have been put to use substantially in analyzing pulsating and cataclysmic variable stars, they have not been applied to study several other astrophysical objects like non-compact binary systems\cite{buchler1995chaotic,cannizzo1988chaos}. The binaries belong to multi-stellar systems that are the most common type of stellar systems, thought to form over 60 $\%$ of all stellar systems in the universe. When the inclination of the system is such that the component stars eclipse each other, leading to a variation of light intensity, the system is called an eclipsing binary or variable \cite{kopal2012dynamics}. Eclipsing binaries are further classified morphologically as detached and close binary star systems. Close binary systems exhibit thermal contact and have the ability to exhibit mass transfer. This offers the prospect of understanding many interesting physical phenomena like stellar mergers, thermal relaxation oscillations, magnetic stellar winds etc\cite{tylenda2011v1309,qian2003overcontact,lee2004period}. After the  observation of a merger event in 2008 in V1309 Scorpii, considerable speculation on future merger events have lead to a deepened interest in these systems \cite{tylenda2011v1309,molnar2017prediction,socia2018kic,kobulnicky2019detecting}. 

Overcontact binaries, called W UMa stars after their prototype, are a subclass of close binary stars which shares a common envelope of gas. They are characterized by both components of the binary star exceeding their Roche lobes, implying that the companion stars are in physical contact with each other \cite{kallrath2009eclipsing}. One quantity used  to characterize the degree of contact in binary systems, is the fill-out factor, $ff$ \cite{kallrath2009eclipsing,wilson1979eccentric}. The fill-out factor is 0 when the component stars in the binary are just in contact and is 1 when they are in complete contact. When both components exceed their Roche lobes, as in the case of overcontact binary stars, mass and energy transfer can occur in either direction \cite{eggleton2006evolutionary,paczynski1967gravitational}. This can lead to changes in periods and stellar mergers leading to red novae\cite{qian2003overcontact,tylenda2011v1309}. While the primary variation in these stars is expected to be due to the orbital motion, many unexplained or partially explained phenomena remain, such as the occurrence of unequal maxima and varying eclipse times\cite{oconnell1951periastron,tran2013anticorrelatedETV}. Various explanations are offered for these, namely the existence of mass transfer, star-spots, presence of a third star, apsidal motion etc \cite{conroy2014kepler}. Many contact binaries are thought to be members of triple systems, which are known to exhibit chaotic behaviour \cite{fabrycky2007shrinking,strogatz2018nonlinear}. 

Apart from the light curves, an important time series used for understanding periodic phenomena in astrophysics is the O-C curve. For eclipsing binaries it is generated by calculating the observed eclipse event minus the predicted time of eclipse. A detailed investigation by \cite{tran2013anticorrelatedETV} on the O-C curves of $32$ contact binaries, suggested that the eclipse time variation may be exhibiting a random walk like behaviour. They also concluded that star-spots might be the most likely cause of eclipse time variations in these stars. Often contact binaries show night to night light variation, suggesting spot evolution at orbital or sub-orbital time scales \cite{csizmadia2006halpha}. This irregular light curve variations at sub-orbital time scales calls for the use of the techniques of nonlinear time series analysis. 

One of the important characteristics of the phase space structure of a chaotic system is its fractal nature or strangeness. The correlation dimension, $D_2$ is an important dynamical measure that is used for measuring this strangeness, and consequently detecting chaos and nonlinearity in the dynamics from time series data. The Grassberger-Procassia algorithm or GP algorithm for calculating $D_2$ is one of the most popular algorithms for calculation of fractal dimensions from time series data embedded into an $M$-dimensional space \cite{gra86}. It has been put to use for detection of chaos from a variety of  datasets like EEG and ECG data, black hole data, photosynthesis data, stock market returns data etc. \cite{buchler1995chaotic,pereda1998eeg, misra2006nonlinear,george2017ecology,scheinkman1989stock}. A saturating non integer $D_2$ value is indicative of deterministic chaos in the underlying dynamics. It may be noted that colored noise data and strange non-chaotic data may also give rise to the above stated condition on $D_2$ \cite{osb89, pra01}. While the former may be eliminated using the method of surrogate data testing, by constraining the power spectrum, the latter needs more thorough investigation using methods such as spectral scaling or bicoherence \cite{the92,george2017detecting}. 

The multifractal ($f(\alpha)$) spectrum is a detailed characterization of the complex fractal structure of the phase space of a dynamical system. Unlike $D_2$ described above which is an average measure, the $f(\alpha)$ spectrum takes the local contributions of different regions into account as well. The range of scales in the $f(\alpha)$ spectrum is a good measure of the underling dynamical complexity and has been put to considerable use in various fields \cite{george2017ecology,harikrishnan2011nonlinear,shekatkar2017ecg}.

Another important nonlinear measure derived from spectral properties is the bicoherence function, which helps to identify quadratic phase coupling between frequencies in a time series \cite{tot08}. We specifically mention the main peak bicoherence function, $b_F(f)$ defined in \cite{george2017detecting}, and used to understand the dynamics of Kepler light curves of RRc Lyrae variable stars \cite{george2017detecting}. 

In this study, we investigate the light curves of $463$ overcontact binary stars in the Kepler field of view \cite{prvsa2011kepler}, using techniques of nonlinear dynamics and search for signs of deterministic chaos by computing their nonlinear measures. The three main quantifiers that we use are correlation dimensions ($D_2$), multifractal measures and main peak bicoherence indices $(b_F(f))$.  All three quantifiers have been put to considerable use to understand various astrophysical phenomena in the past \cite{george2017detecting,odekon2006correlation,misra2004chaotic,harikrishnan2011nonlinear,maccarone2002Higher}. We start by the phase space reconstruction from data and present the computation of $D_2$ and $f(\alpha)$ from the phase space structure. We notice evidence of deterministic chaos in them that explains the extra frequencies, other than the eclipsing frequency and its harmonics, seen in the power spectra of these stars. 

We also study the time series of eclipse time variations, commonly called the O-C (Observed minus Calculated) curves, for four of these stars\cite{sterken2005oc}. The O-C curve is a time series of the observed eclipse event minus the predicted time of eclipse, and hence is a measure of eclipse time variations. Studying variations of the period of a nearly periodic phenomenon is popular across multiple fields, like in heart rate variability in cardiac dynamics, flowering time variation and population dynamics in ecology etc \cite{tejera2010unexpected,Jochner2016,esper20061200}. Many features in these variability curves were explained due to nonlinearity \cite{wu2009chaotic,iler2013nonlinear,iyengar2016co}. Previous studies have noted that the O-C curve variations for overcontact binaries are random walk like\cite{tran2013anticorrelatedETV}. We see that the variation of the timings of maxima in overcontact binary stars is similar to such variations in chaotic systems. This finding suggests that the variation in eclipse times may have deterministic origin.

We further confirm our findings by studying the bicoherence from the spectra of all the stars. We see that the bicoherence between the orbiting frequency and other frequencies is significant in many of the stars, suggesting that the orbiting frequency is quadratically coupled to frequencies arising due to other mechanisms in the system. 

Finally, we check for correlations that may exist between the nonlinear characteristics of these binary systems and their degrees of contact quantified using the fill-out factor mentioned above\cite{kallrath2009eclipsing}. By considering a subset with restrictions on spectral class and mass ratio, we see that the correlations become more pronounced. We also consider the correlations that exists between many of the other relevant parameters of the binary star system, namely period, effective temperature and mass ratio, with the calculated nonlinear parameters, namely $D_2$, the multifractal measures and the $b_F(f)$. We argue that the existence of significant correlations, between the nonlinear measures and the fill-out factor, imply that the fractal properties and complexity of the stars may be changing as the binary star systems evolve over time.

\section{Embedding and Fractal Measures} \label{sec:frac}
The dataset used in the study consists of the light curves of all the eclipsing binaries listed as overcontact binary stars in the second revision of the Kepler eclipsing binary catalog, totaling to 463 stars \cite{prvsa2011kepler}. One of the primary problems associated with Kepler datasets has been the presence of gaps in the light curve. In a series of two papers, we have previously identified a tolerable range of gap sizes and frequencies, within which reliable conclusions can be drawn from the $D_2$ and $f(\alpha)$ curves\cite{george2017ecology,geo15}. The gap ranges in the Kepler dataset fall well within the identified gap ranges. Typical light curves and power spectra\footnote{For power spectra calculations(and subsequently bicoherence), we first divide the light curves into $k$ evenly sampled segments. The Fourier transforms are calculated individually over each of the segments and averaged over different segments to yield an average value.} for four overcontact binaries taken from this set are shown in Figure \ref{fig:lcpsbin}. We see that the power spectra show peaks with considerable power at half integer positions, characteristic of period doubling in their dynamics \cite{hilborn2000chaos}. A series of such repeated period doublings is a characteristic route that a nonlinear dynamical system takes to reach chaos. In terms of the power spectrum, a period doubled limit cycle shows peaks at the original time period and its harmonics and smaller peaks at half of the original period and its harmonics. This period doubling has been recently reported as evidence of nonlinear phenomena in many astrophysical objects \cite{szabo2010does,moskalik2015kepler,plachy2016first}. The period doubling is characteristic of the inherent nonlinearity present in the system leading to chaotic behaviour. This motivates the use of nonlinear time series analysis tools and fractal measures to understand their dynamics. 

We start the analysis by reconstructing the phase space, based on Taken's theorem, from the observational data \cite{takens1981detecting}. From the reconstructed phase space structure, we calculate the correlation dimension. Using the modified GP algorithm proposed in \cite{har06}. The algorithm first takes the uniform deviate of the light curves by converting the amplitude distribution of the light curves to a uniform distribution \cite{press2007numerical}. This helps to eliminate any differences between the light curves as a result of their differing amplitude distributions. The point where the auto-correlation function falls to $\frac{1}{e}$ is taken as the delay time, $\tau$, which is used to construct delay vectors. If $I(t)$ is the light curve and $I_u(t)$ is its uniform deviate, a delay vector in an M dimensional space, at a time $t_i$, would be constructed as
\begin{equation}
    \Vec{v}_i=[I_u(t_i), I_u(t_i+\tau), ..., I_u(t_i+M\tau)]
\end{equation}
The reconstructed phase space projections for the four typical light curves considered in Figure \ref{fig:lcpsbin} are shown in the upper panel of Figure \ref{fig:phspd2}.

\subsection{Correlation Dimension}\label{sec:d2}
For the reconstructed phase space trajectory of each star, we count the relative number of vectors within an $M$-cube of length R of each vector, labeled by $p(R)$, for $N_c$ chosen centers and average this about these selected centers, to get the correlation sum, $C_M$.

\begin{figure}[t]
\begin{center}
\includegraphics[height=.6\textheight]{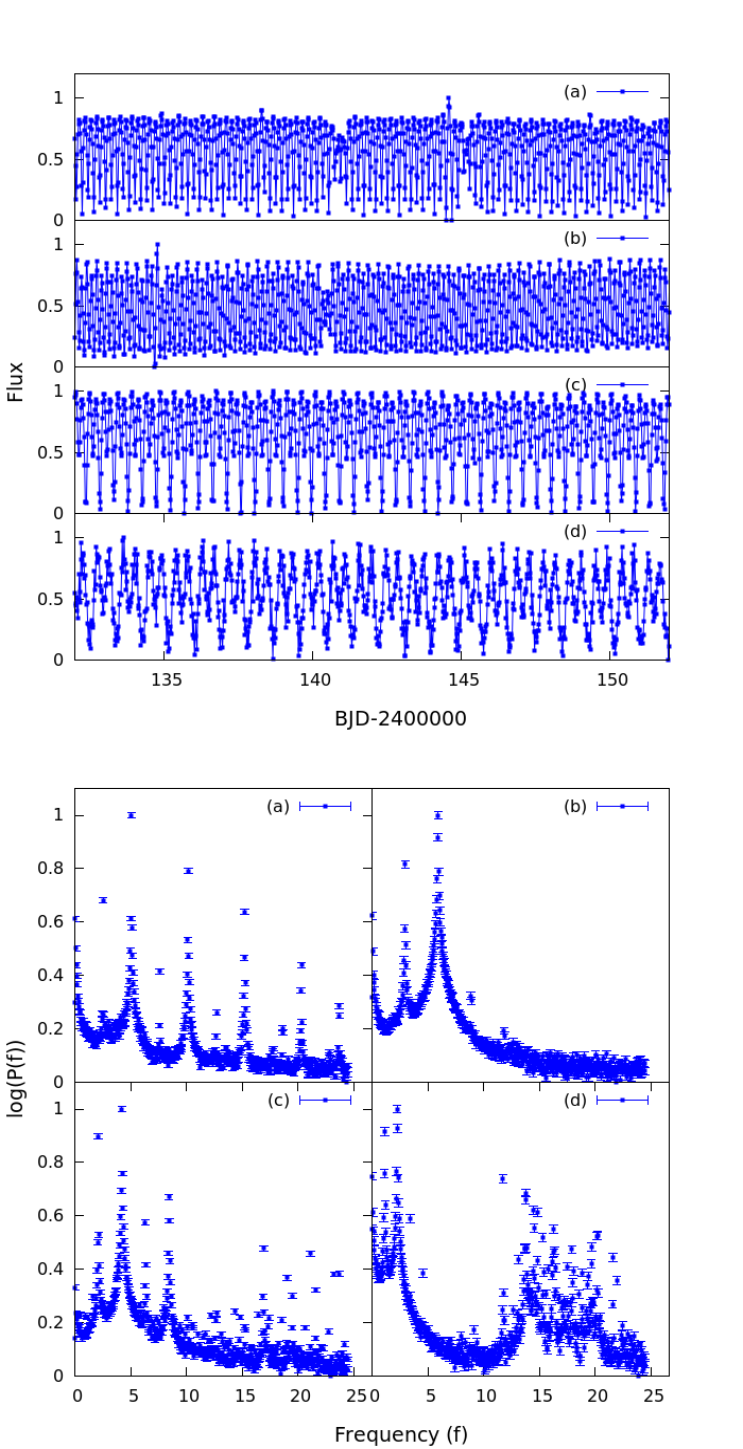}
\end{center}
\caption{\label{fig:lcpsbin}Light curves and power spectra of four typical overcontact binary stars. The upper panel corresponds to the light curves and the lower panel shows the corresponding power spectra. The stars under consideration are (a)KIC 4909422, (b)KIC 6368316, (c)KIC 7657914, and (d)KIC 8800998. All four power spectra show peaks with considerable power at half-integer multiples of the primary peak, characteristic of period doubling. Both the light curve flux and the power spectra have been re-scaled to the range [0:1] for clarity.}
\end{figure}

\begin{figure}[t]
\begin{center}
\includegraphics[height=.6\textheight]{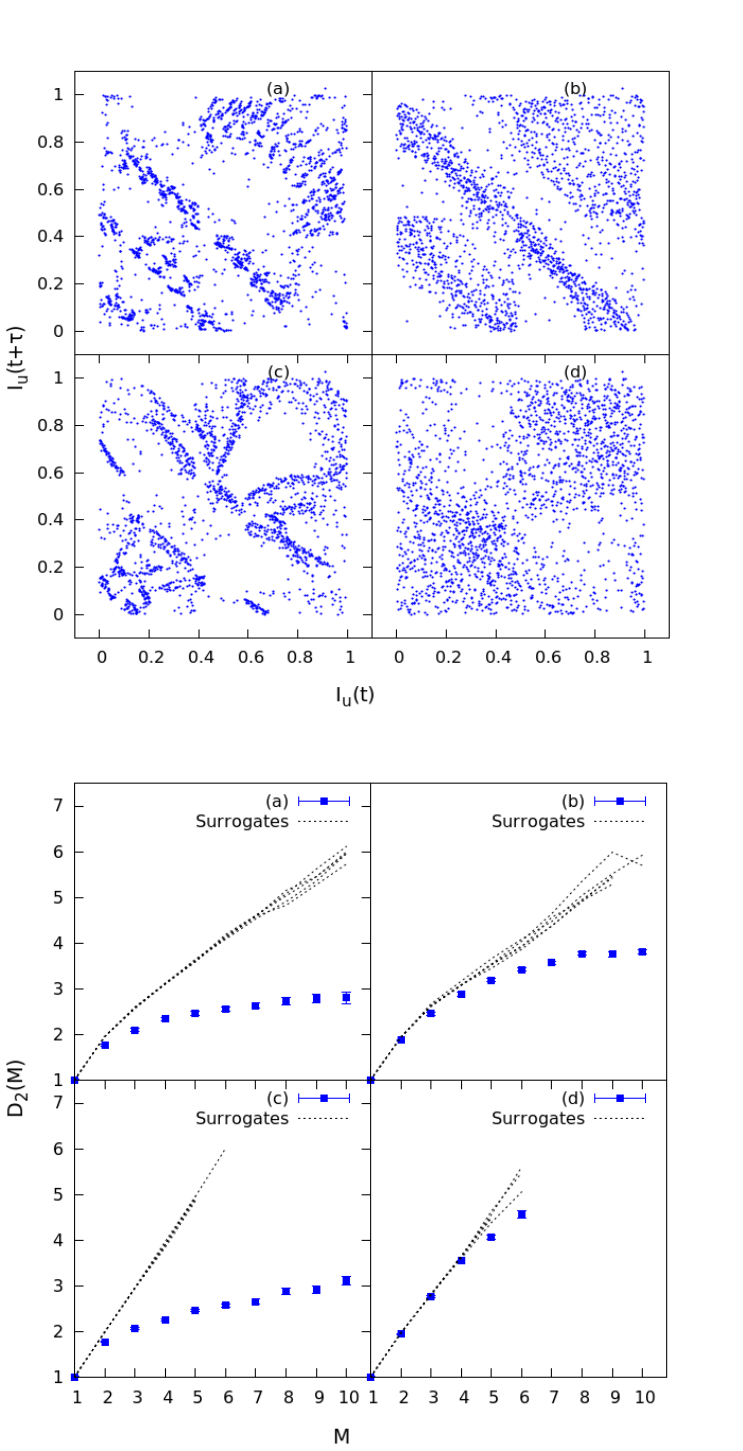}
\end{center}
\caption{\label{fig:phspd2} 2-D projections of the embedded phase space structure of four typical overcontact binary stars in Figure \ref{fig:lcpsbin} are displayed in the upper panel. $D_2(M)$ vs $M$ plots corresponding to these light curves and five of their surrogates are shown in the lower panel. (a), (b) and (c) shows saturation away from surrogates with M, while (d) does not show any significant saturation.}
\end{figure}

\begin{equation}
    C_M(R)=\frac{1}{N_c}\sum_{i}^{N_c}p_i(R)
\label{eq:corrsum}
\end{equation}
The correlation sum, $C_M$ scales with R as 
\begin{equation}
    C_M(R) \approx R^{D_2}
\end{equation}

where $D_2$ is the correlation dimension. Thus $D_2$, can be obtained as the slope from the logarithmic plot of $C_M(R)$ vs R. In general $D_2$ is calculated for increasing embedding dimension, $M$. Then $D_2(M)$ vs $M$ curve is fitted using the function, 
\begin{eqnarray} 
f(M)\; & = & \;\Big {(}{D_2^{sat} - 1 \over M_d -1}\Big {)} (M-1) +1 
\;\;\;\; \hbox {for} \;\;\;\; M < M_d \nonumber \\ 
          & = & \; D_2^{sat} \;\;\;\; \hbox {for} \;\;\;\; M 
\geq M_d 
\label{fit} 
\end{eqnarray} 
which gives the saturated value of the correlation dimension,$D_2^{sat}$\cite{har06}.

We calculate this $D_2^{sat}$ (referred to simply as $D_2$ hereafter) for all the stars in our dataset. The presence of a saturating $D_2$ is indicative of the underlying deterministic chaos. However, as mentioned earlier, colored noise processes may lead to the generation of a random fractal curve which can lead to saturation of the $D_2$ vs M curve \cite{osb89}. We can differentiate between the two using the method of surrogate analysis. For this, we generate surrogates by the method of Fourier phase randomization, implemented using the IAAFT algorithm in the TISEAN package \cite{sch00, heg99tisean}. We consider five such surrogate datasets for every light curve. $D_2$ values for the surrogate datasets are compared with the $D_2$ for the original datasets. The deviation of $D_2$ of data from that of surrogates indicates that the observed saturation is due to deterministic chaos. We measure this deviation, using the $nmsd$ (normalised mean sigma deviation) measure defined as \cite{har06}
\begin{equation}
nmsd^2   = \frac{1}{M_{max} -1} \sum_{M = 2}^{M_{max}} \Big {(}\frac{D_2 (M) - < D_2^{surr} (M) >}{\sigma^{surr}_{SD} (M) }\Big {)}^2 
\label{eq:nmsd}
\end{equation}

\begin{figure}
\begin{center}

\includegraphics[width=.6\columnwidth]{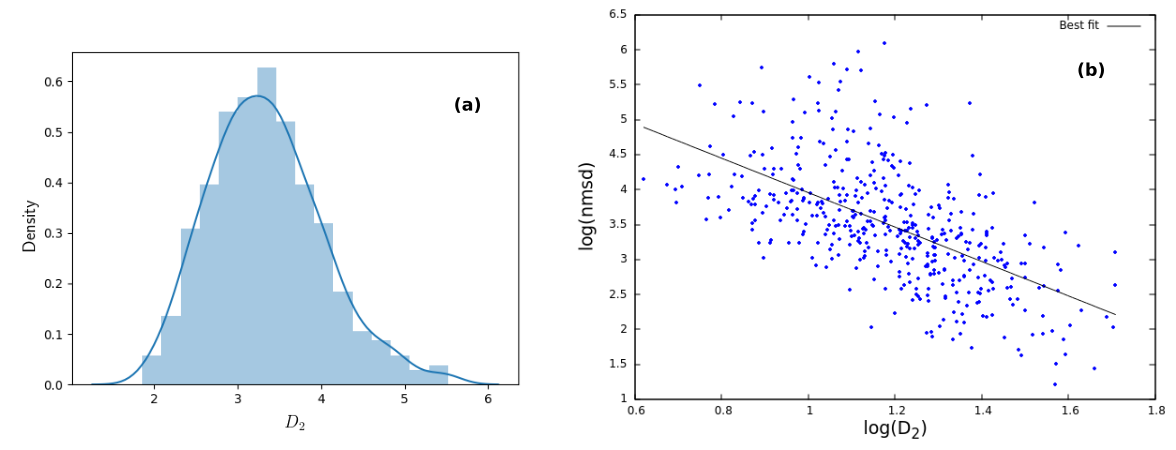}

\end{center}
\caption{\label{fig:d2hist_nmsd} (a) Normalized histogram and kernel density estimate of $D_2$ for all the light curves considered. (b)Variation of the log of the $nmsd$ with log of $D_2$. We see a fall in $nmsd$ as $D_2$ increases. Hence the increase $D_2$ may be attributed to an increase in stochasticity or a change in the underlying equations of the system.}
\end{figure}

The $D_2(M)$ vs $M$ curves for the four light curves considered in Figure \ref{fig:lcpsbin} above are shown in the lower panel of Figure \ref{fig:phspd2}, along with $5$ surrogate datasets. For the datasets we have considered, we see that almost all the light curves show significant deviation of the $D_2$ of  data from that of surrogates, implying that a large majority of overcontact binaries show deterministic and nonlinear behavior. The distribution of the $D_2$ values for all the light curves considered in this study is shown in Figure \ref{fig:d2hist_nmsd}a. We find that there cases with low $nmsd$ values and they correspond to larger values of $D_2$, as is clear from Figure \ref{fig:d2hist_nmsd}b. It is known that noise contamination is one of the reasons for a decrease in $nmsd$ values in a dynamical system. Hence the binaries which show lower $nmsd$ values may be having stochastic factors affecting their dynamics. We will discuss this further in the light of the analysis presented in subsequent sections.

\subsection{Eclipse time variations}
\begin{figure}
\begin{center}
\includegraphics[width=.8\columnwidth]{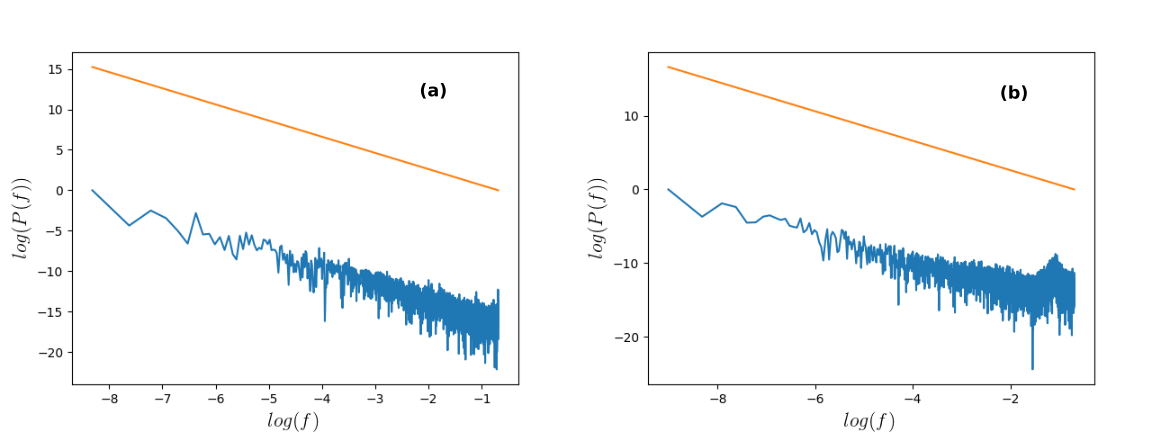}
\end{center}
\caption{\label{fig:ps_etv}Power spectra for the eclipse time variation for (a)R{\"o}ssler system and (b)KIC 4909422. The $\frac{1}{f^2}$ line is shown in both cases for comparison. }
\end{figure}

We present the $D_2$ analysis for O-C (Observed minus Calculated) curves for the four sample stars considered in Figure \ref{fig:lcpsbin}, along with $D_2$ for five surrogate datasets. We compare the results with the variations in timings for maxima for two standard nonlinear systems, the R{\"o}ssler system and a forced damped pendulum in the chaotic regime. The O-C curves for all the time series are generated using the following formula\cite{borkovits2015eclipse}
\begin{equation}
\Delta=T_i-T_0-i \times P^s
\label{eq:occurve}
\end{equation}
where, $T_i$ is the observed time of the $i^{th}$ maxima, $T_0$ is the initial time of observation, and $P_0$ is the mean period of maxima. We calculate $P_0$ as the average of time interval between successive maxima in the time series. We observe random walk like features in the time series of maximum variation in deterministic chaotic systems. We illustrate this by plotting the power spectrum for the R{\"o}ssler system and for the star KIC 4909422(Figure\ref{fig:ps_etv}), both of which show $\frac{1}{f^2}$ like behavior. The values for $D_2$ and $nmsd$ for the four O-C curves and two deterministic dynamical systems considered is shown in Table \ref{table:occurve}. The significant values of $nmsd$ indicate that the eclipse time variations may also have some underlying nonlinear dynamical behavior. 

\begin{table*}[ht] 
\caption{$D_2$ and $nmsd$ values for the O-C curves of the four overcontact stars shown in Figure \ref{fig:lcpsbin}, and two chaotic dynamical systems.} % title of Table 
\centering % used for centering table 
\begin{tabular*}{.5\textwidth}{@{\extracolsep{\fill}}c c c} % centered columns (3 columns) 
\hline\hline %inserts double horizontal lines 
Kepler ID  & $D_2$  & $nmsd$ \\ [0.5ex] % inserts table %heading 
\hline % inserts single horizontal line 
$4909422$ & $1.74$ & $2.77$ \\ 
$6368316$ & $1.50$ & $3.56$ \\ 
$7657914$ & $1.92$ & $7.42$ \\
$8800998$ & $1.74$ & $2.05$\\
$R${\"o}$ssler$ & $3.84$ & $1.63$ \\
$Pendulum$ & $3.39$ & $3.33$\\
%[1ex] % [1ex] adds vertical space 
\hline %inserts single line 
\end{tabular*} 
\label{table:occurve} % is used to refer this table in the text 
\end{table*}

\subsection{Multifractal Spectrum}\label{sec:mfs}

The reconstructed phase-space or attractor for most of the nonlinear systems has a multifractal structure which is characterized by a set of generalized dimensions $D_q$. which provides a measure of the non-uniformities in the distribution of points in the attractor. 
For computing the generalized dimensions $D_q$ , we define a generalized correlation sum as\cite{hilborn2000chaos}
\begin{equation}
    C_q(R)=\sum_{j=1}^{N}p_j^{q}
    \label{eq:gencorrsum}
\end{equation}
The generalised dimension $D_q$ is then defined as
\begin{equation}
    D_q=\lim_{R\to0}\frac{1}{q-1}\frac{lnC_q(R)}{lnR}
    \label{eq:GenDimension}
\end{equation}

There is a parallel approach in which we cover the attractor with boxes of size $R$. The probability $p_i$ of points inside the $i^{th}$ box scales as
\begin{equation}
    p_i(R)=R^{\alpha_i(R)}
\end{equation}
The number of boxes with $\alpha$ between $\alpha$ and $\alpha+\Delta\alpha$ is given by $n(\alpha)$ \cite{jensen1985global}, which relates to the size of the box $R$ as,
\begin{equation}
    n(\alpha,R) \propto R^{-f(\alpha)}
\end{equation}

These two characterizations  using  $(D_q,q)$ and $(f(\alpha),\alpha)$, are related to each other through a Legendre transform \cite{jensen1985global,hilborn2000chaos}.

\begin{equation}
    \alpha=\frac{d}{dq}[(q-1)D_q]
\end{equation}

\begin{equation}
    f(\alpha)=q\alpha - (q-1)D_q    
\end{equation}

The computational method used to compute multifractal measures uses above relation to get the $f(\alpha)$  from $D_q$ \cite{harikrishnan2009computing}.
The $f(\alpha)$ curve thus obtained is characterised by four parameters $\alpha_{min}$, $\alpha_{max}$, $\gamma_1$ and $\gamma_2$ as
\begin{equation}
f(\alpha)=A(\alpha-\alpha_{min})^{\gamma_1} (\alpha_{max}-\alpha)^{\gamma_2}
\label{eqn:fal}
\end{equation}

Since for more than $80\%$  of the lights curves used in the study, the $D_2$ values lie below $4$ (Figure \ref{fig:d2hist_nmsd}), and hence $M$ is chosen to be $4$ for our $f(\alpha)$ calculations. 

\begin{figure}[h]
\begin{center}
\includegraphics[width=.6\columnwidth]{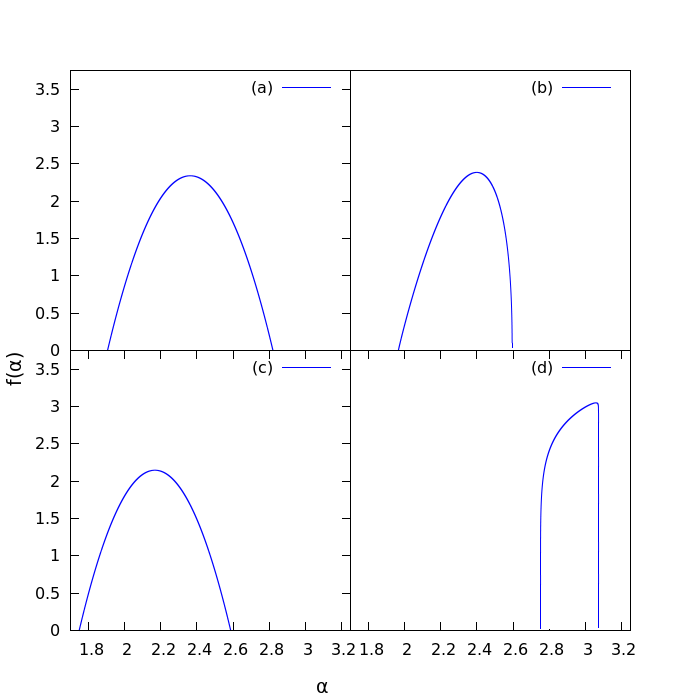}
\end{center}
\caption{\label{fig:A1} $f(\alpha)$ vs $\alpha$ plots for the 4 sample eclipsing binary stars considered. The figures correspond to (a)KIC 4909422, (b)KIC 6368316, (c)KIC 7657914, and (d)KIC 8800998. The narrow spectrum corresponding to (d) is indicative of noisy behavior, as suggested by Figure \ref{fig:phspd2}.}
\end{figure}

We calculate, the $\alpha_{min}$ and $\alpha_{max}$ for all the light curves considered and the difference $\alpha_{max}- alpha_{min}$ is a measure of the complexity of the phase space structure. The $f(\alpha)$ curves of the four sample overcontact binary stars are shown in Figure \ref{fig:A1}. For the data of light curves used, we find the numerical error is more for calculation of $\alpha_{max}$.

\section{Main Peak Bicoherence} \label{sec:mpb}
We supplement the studies on fractal measures with studies using bicoherence computed from the Fourier transforms of the light curves. The bicoherence function measure that quantified the extent of the quadratic coupling between the different frequencies in the power spectrum of the system. 
It is defined as 
\begin{equation}
B(f_1,f_2)=\frac{|\sum\limits_{i=1}^k A_i(f_1) A_i(f_2) A_i^*(f_1+f_2)|}{\sum\limits_{i=1}^k|A_i(f_1) A_i(f_2) A_i^*(f_1+f_2)|}
\end{equation}
where $A(f)$ is the Fourier transform of the signal at $f$ and $A^*(f)$ is the conjugate of the Fourier transform. The bicoherence essentially checks if the Fourier component at frequencies $f_1$ and $f_2$ are related to the component at $f_1+f_2$. Since such a relation is expected from a system that has a quadratic response,  for a process where there is no phase relation between the pairs, the bicoherence would fall with number of segments, $k$, as $\sqrt{\frac{1}{k}}$ similar to a 2 dimensional random walk. The bicoherence function offers many inherent advantages over the power spectrum, since it retains the phase relation between frequency pairs. The bicoherence function of various chaotic systems has been studied in the past\cite{pez90,cha93,elg93}, but has not been put to use to quantify nonlinearity and chaos from real world time series until recently\cite{george2017detecting}.
The main peak bicoherence function (b$_F$(f)) is defined by replacing one of the frequencies, $f_1$ with the maximal power spectral peak, $F$ \cite{george2017detecting}.
\begin{equation}
b_{F}(f)=\frac{|\sum\limits_{i=1}^k A_i(F) A_i(f) A_i^*(F+f)|}{\sum\limits_{i=1}^k|A_i(F) A_i(f) A_i^*(F+f)|}
\end{equation}
For chaotic systems, it is often sufficient to consider this main peak function, instead of considering the entire plane\cite{george2017detecting}.
\begin{figure}[h]
\begin{center}
\includegraphics[width=.6\columnwidth]{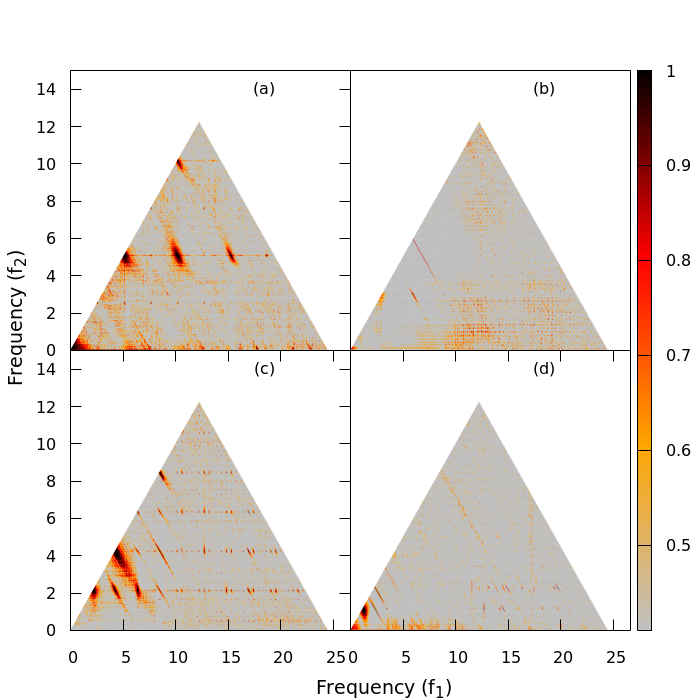}
\end{center}
\caption{\label{fig:fbc} Full bicoherence plots for the four typical eclipsing binary stars. We see very few frequency pairs with significant bicoherence in (b) and (d), whereas (a) and (c) show significant bicoherence for many frequency pairs.}
\end{figure} 
As a first step, since Kepler data has gaps in observation, we first construct $N$ evenly sampled segments of $1024$ points each. The Fourier transforms are calculated, using the FFT algorithm, for the individual evenly sampled segments and averaged to get an estimate of the bicoherence. The full bicoherence plots calculated for the 4 stars considered previously is shown in Figure \ref{fig:fbc}. The maximal peak, $F$, in the power spectrum corresponds to the eclipsing frequency of the stars. This analysis is important as it examines whether the eclipsing frequency is coupled the other frequencies present in the system. The $b_F(f)$ graphs for the 4 stars considered is shown in Figure \ref{fig:mpb}.

\begin{figure}
\begin{center}
\includegraphics[width=.6\columnwidth]{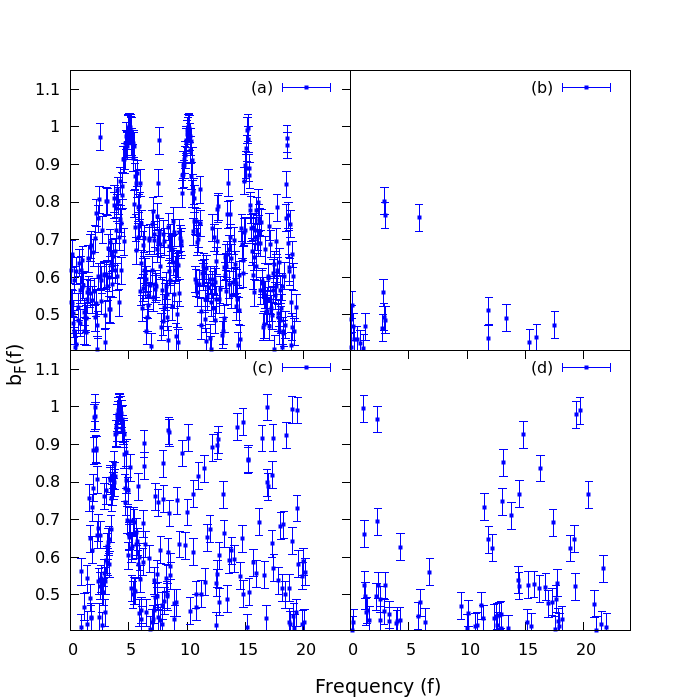}
\end{center}
\caption{\label{fig:mpb} $b_F(f)$ vs $f$ plots for the four typical eclipsing binary stars considered above. Only bicoherence values above $99\%$ significance are plotted. As in the case of Figure \ref{fig:fbc}, (a) and (c), show significant coupling with the eclipsing frequency whereas (b) and (d) have much less frequencies that show significant value for bicoherence.  The error bar on the bicoherence for $N$ segments is given by $\frac{1}{N}$ \cite{maccarone2002Higher}.}
\end{figure}

\begin{figure}
\begin{center}
\includegraphics[width=.6\columnwidth]{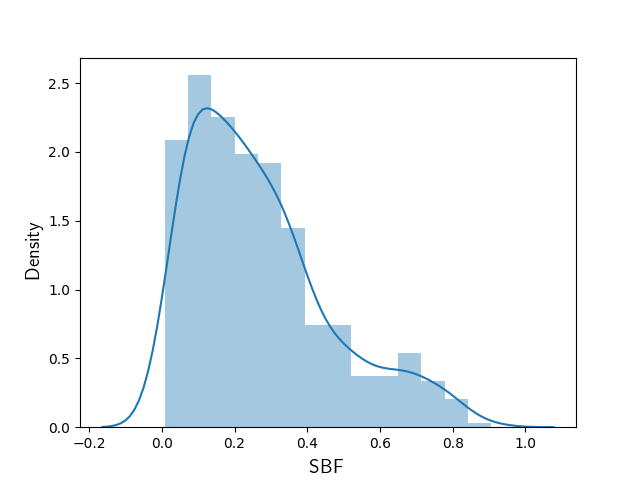}
\end{center}
\caption{\label{fig:smbc} Normalized histogram and kernel density estimate of Significant Bicoherence Fraction (SBF) for all the light curves considered.}
\end{figure}

We measure the relevance of the bicoherence estimated through summation over $N$ segments of the time series, through a significance measure. The $99\%$ significance threshold for $N$ segments of the time series is $\sqrt{\frac{9.2}{2N}}$\cite{cha93}. We then calculate the fraction of frequencies which have a bicoherence value above this threshold, and term it the significant bicoherence fraction ($SBF$). The significant bicoherence fraction is a measure of the extend of coupling of different frequency components in the Fourier spectrum with the primary or eclipsing frequency. It is calculated for all the stars considered. The distribution of the significant fraction for $b_F(f)$ is shown in Figure \ref{fig:smbc}. We see that a comnsiderable number of stars have a large $SBF$. This implies that the eclipsing frequency is quadratically coupled with the other frequencies in these stars. Another set of stars however show little or no coupling with the eclipsing frequency. This indicates that either these systems are more stochastic in nature or that the dominant nonlinearity in them is no longer of quadratic order. 

\section{Correlations between binary parameters and nonlinear measures} \label{sec:corr}
In the previous section we establish the presence of deterministic nonlinearity and quantify it using $D_2$, $f(\alpha)$ and $b_F(f)$. In this section we investigate whether these quantifiers are related to the physical properties of the stars themselves. We note that changes in the parameters of a nonlinear dynamical system can result in changes in the nonlinear quantifiers of that system. Hence we look into the correlations between the nonlinear characterizers of the light curve calculated in the previous sections, and the physical properties of the overcontact binary stars. 

\subsection{Correlations with degree of contact}
The fill-out factor or contact parameter, $ff$ , is a measure of the degree of contact between the companions of the binary. It is defined using $\Omega$, the surface of the common envelope, $\Omega^I$, the potential at the inner Lagrangian surface and $\Omega^O$ at the outer Lagrangian surface, as 

\cite{kallrath2009eclipsing,wilson1979eccentric}
\begin{equation}
    ff=\frac{\Omega^I-\Omega}{\Omega^I-\Omega^O}
\end{equation}
The form given by the above equation is such that when the stars are just in contact (when the potential at the surface of the star equals the
potential at the inner Lagrangian surface) $ff=0$, whereas when they are in complete contact (when the potential at the surface of the star equals the potential at the outer Lagrangian surface), $ff$ becomes 1. The $ff$ values of binaries used in the study are extracted from \cite{prvsa2011kepler}\footnote{Data available at http://keplerebs.villanova.edu/v2}.

\begin{figure}
\centering
\includegraphics[width=.8\columnwidth]{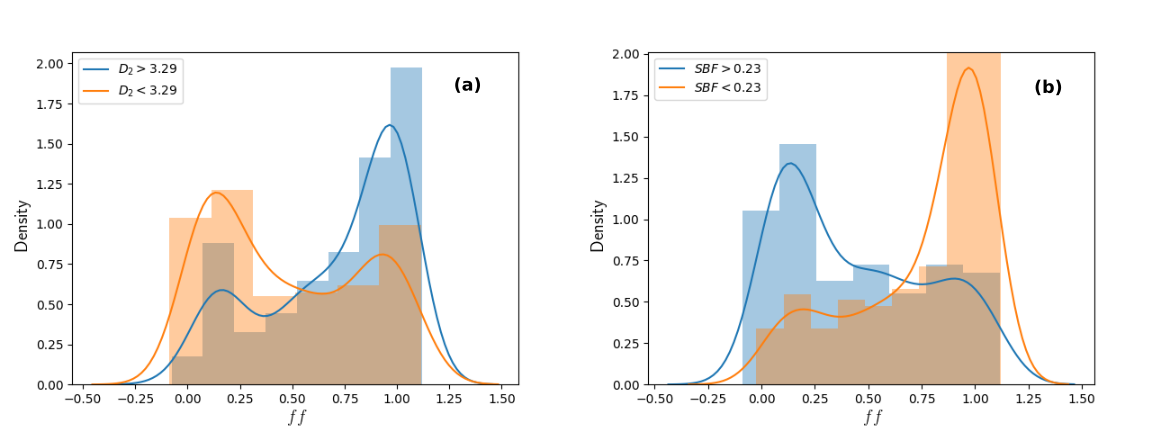}

\caption{\label{fig:kdeff} Plots of kernel density estimates of fill-out factors for (a)$D_2 < 3.29$ and $D_2 > 3.29$ (b)$SBF < 0.23$ and $SBF > 0.23$. One can see two different distributions when we subdivide the parameters into two. In (a) the skews of the distributions corresponding $D_2 < D_2^{med}$ and $D_2 > D_2^{med}$, towards higher $ff$ and lower $ff$ respectively, suggests that higher $D_2$ implies a more evolved system with higher $ff$ values. Similarly in (b) we see that the stars start to loose coupling with the eclipsing frequency as it evolves. } 
\end{figure}

To check for correlations of $ff$ with computed $D_2$ and $SBF$ values, we first construct two separate subcategories of binary stars corresponding to high and low ranges of $D_2$ and $SBF$. We correspondingly then consider the distribution of $ff$ for the two categories. The ranges of $D_2$ and $SBF$ are decided using the medians of the $D_2$ and $SBF$ distributions. The values for $D_2^{med}$ is $3.29$ and for $SBF^{med}$ is $0.23$. We take the cases where $D_2>3.29$ with $226$ samples and $D_2<3.29$ with $227$ samples (Figure \ref{fig:kdeff}a). For the bicoherence we consider the $SBF$ above and below $0.23$ (Figure \ref{fig:kdeff}b). The two subcategories seem to possess distributions that are peaked at different values of $ff$. Hence we proceed to examine the correlations between them. In all our analysis of correlations, we use the Spearman rank-order correlation coefficient, $\rho_S$, to describe the extend of correlation, since it checks for any monotonic increase \cite{press2007numerical}.  

\begin{figure}
\centering
\includegraphics[width=.6\columnwidth]{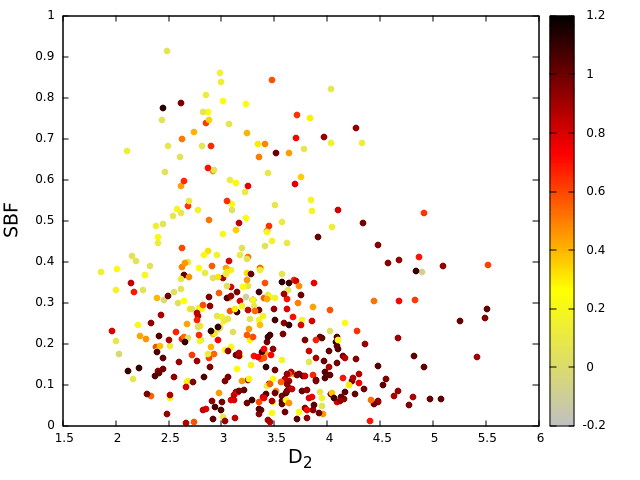}

\caption{\label{fig:d2vssbfvsff} Scatter plot of all the overcontact stars as a function of $D_2$ and $SBF$. The color code shows the $ff$. We see that the top left of the graph corresponding to higher $SBF$ and lower $D_2$, shows a lower $ff$ whereas the bottom right corresponding to higher $D_2$ and lower $SBF$ shows a higher value for $ff$. } 
\end{figure}

We see significant correlation, between $D_2$ and $ff$ ($\rho_S=.33$, p-value of $6.9 \times 10^{-13}$) and between $SBF$ and $ff$ ($\rho_S = -0.44$, p-value of $2.2 \times 10^{-23}$) for all the stars considered. The very low p-value indicates that the probability that the correlation appeared spuriously is negligible. Hence, high $ff$ corresponds to a higher $D_2$ and a lower $SBF$\footnote{It may be noted that the correlation between $D_2$ and $SBF$ is much smaller at $\rho_S = -0.24$, suggesting that these correlations with $ff$ do not follow from each other}. We illustrate these observations in Figure \ref{fig:d2vssbfvsff} and show that $ff$ is directly correlated to $D_2$ and inversely to $SBF$. Astrophysically the $ff$ is a measure of the extend of contact between the component stars. Hence greater contact seems to imply larger phase space dimension and lower coupling with the orbiting frequency. 

\begin{figure}
\centering

\includegraphics[width=.8\columnwidth]{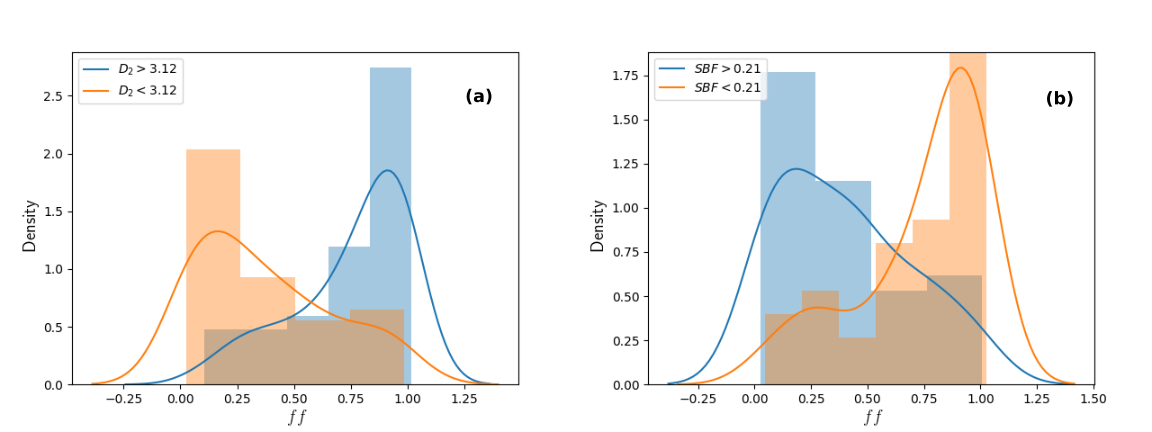}

\caption{\label{fig:ResD2vsFF} Plots of kernel density estimates of fill-out factors for (a)$D_2 > D_2^{med}$ and $D_2 < D_2^{med}$($D_2^{med}=3.12$) and (b)$SBF>SBF^{med}$ and $SBF<SBF^{med}$($SBF^{med}=0.21$) for $\frac{1}{2} < q< 2$; $6000 < T_{eff}< 7000$. The significantly different distributions in these plots suggests that as one goes into more restricted sub-populations, the dimension of the system and the coupling of frequencies in the system with the eclipsing frequency become more closely linked to the fill-out factor.} 
\end{figure}

To check the effect of the evolution of stars on nonlinear measures we confine our analysis to restricted ranges in mass ratio and temperature. Within this range, any change in the dynamics of the system can be thought to be due to change in $ff$ alone. We restrict the mass ratio as $\frac{1}{2}<q<2$ and the effective temperature as $6000 < T_{eff} < 7000$. We get a correlation between $D_2$ and $ff$ as $\rho_S = 0.69$ and p-value of $5.2\times10^{-14}$ , for a sample size of $91$ stars \cite{kirk2016kepler}. The correlation between $SBF$ and $ff$ too increases to $\rho_S = -0.60 $ for this subclass of stars with a p-value of $2.6 \times 10^{-10}$. The kernel density plot of $ff$ in different ranges of $D_2$ and $SBF$ in this subcategory is shown in Figure \ref{fig:ResD2vsFF}. Among the $f(\alpha)$ parameters, a significant correlation can be seen only between $\alpha_{min}$ and $ff$, with $\rho_S=0.43$ (p-value $1.8\times10^{-5}$). In the astrophysics parlance, the temperature range considered roughly corresponds to the F spectral class. Contact binaries from these earlier spectral types are thought to form a different subclass of overcontact binaries, called the A-type W UMa stars\cite{skelton2009modelling}. A list of ten stars with the values of their parameters and their calculated measures from this restricted sub-population is shown in Table \ref{table:EgTable}. 

\begin{table*}[ht] 
\caption{Intrinsic and calculated nonlinear parameters for a set of $10$ sample stars which have $6000 < T_{eff} < 7000$ and $0.5<q<2$.} % title of Table 
\centering % used for centering table 
\begin{tabular*}{\textwidth}{@{\extracolsep{\fill}}c c c c c c c c c c c} % centered columns (6 columns) 
\hline\hline %inserts double horizontal lines 
$KID$  & $Period$  & $q$ & $T_{eff}$ & $ff$  & $D_2$ & $SBF$ & $\alpha_{min}$ & $\alpha_{max}$ & $\gamma_{1}$ & $\gamma_{2}$\\ [0.5ex] % inserts table %heading 
\hline % inserts single horizontal line 

$3437800$ & $0.36$ & $0.91$ & $6185$ & $0.70$ & $4.40$ & $0.02$ &$2.59$& $1.90$ & $0.38$ & $0.14$\\ 
$4273411$ & $1.22$ & $1.36$ & $6975$ & $1.01$ & $4.26$ & $0.08$ &$1.94$ & $1.77$ & $0.21$ & $0.03$\\ 
$5198934$ & $0.83$ & $0.89$ & $6905$ & $0.84$& $3.12$  &$0.02$ & $1.72$ & $1.71$ & $0.95$ & $0.75$\\ 
$7339123$ & $0.35$ & $1.18$ & $6138$ & $0.16$ & $2.98$ & $0.23$ & $2.54$ & $1.71$ & $1.0$ & $0.99$\\ 
$9002076$& $0.48$ & $1.50$ & $6434$ & $1.01$ & $3.76$ & $0.10$ & $2.65$ & $2.40$ & $0.01$ & $0.05$\\ 
$9108579$ & $1.17$ & $1.45$ & $6386$ & $0.99$ & $3.52$ & $0.67$ & $2.75$ & $1.75$ & $1.0$ & $0.40$\\
$10680475$ & $0.35$ & $1.19$ & $6082$ & $0.30$ & $3.05$ & $0.38$ & $2.34$ & $1.58$ & $1.0$ & $1.0$\\ 
$10877703$ & $0.44$ & $0.83$ & $6106$ & $0.10$ & $3.17$ & $0.39$ & $2.31$ & $1.56$ & $1.0$ & $0.99$\\
$11154110$ & $0.53$ & $0.77$ & $6938$ & $0.05$ & $2.61$ & $0.66$ & $2.71$ & $2.17$ & $0.44$ & $0.13$\\  
$11460346$ & $0.39$ & $1.19$ & $6273$ & $0.93$& $4.67$ &$0.08$ & $2.82$ & $2.10$ & $0.19$ & $0.04$\\ 
%[1ex] % [1ex] adds vertical space 
\hline %inserts single line 
\end{tabular*} 
\label{table:EgTable} % is used to refer this table in the text 
\end{table*}

\begin{figure}
\begin{center}
\includegraphics[width=.6\columnwidth]{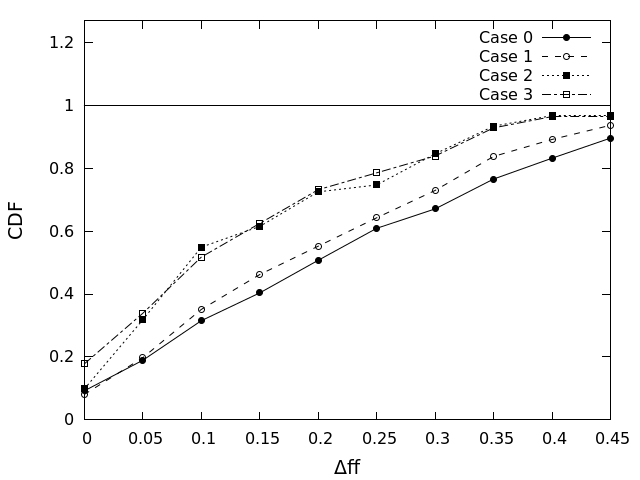}
\end{center}
\caption{\label{fig:cdf} Cumulative density function of the deviations of $ff$ from the best fit line for (a) Case 0 : No restrictions on parameters, (b) Case 1 : $\frac{1}{2} < q< 2$ (c) Case 2 : $\frac{1}{2} < q< 2$, $6000 < T_{eff}< 7000$ and (d)Case 3: $\frac{2}{3} < q< \frac{3}{2}$, $6000 < T_{eff}< 7000$. (a) and (d) suggests that, using simple linear regression, we can find a value of $ff$ in restricted sub-populations of binary stars, with reasonable accuracy.}
\end{figure}

We observe that for the restricted sub-population the correlation dimension of the system evolves as the system evolves in a more pronounced manner as compared to the whole population. Similar to the period-mass correlations derived in \cite{gazeas2008angular}, this understanding can provide an estimate of $ff$ for an unknown binary from its light curve through the calculation of $D_2$ and $SBF$. To illustrate this point, we consider a linear regression of $SBF$ and $D_2$ to get an approximate value of $ff$. In this analysis we derive a linear relation between $ff$, $D_2$ and $SBF$. The difference in $ff$ predicted using this linear relation from the $ff$ derived in \cite{prvsa2011kepler} is plotted as a cumulative distribution in Figure \ref{fig:cdf}. It is interesting to note that fill-out factors of over $70\%$ of stars can be predicted within an accuracy of $0.2$ with just a linear regression using $D_2$ and $SBF$, in some of the subcategories considered,.

\section{Results and Discussion}
\label{sec:ResDis}
We analyze the light curves of all the overcontact binary stars in the Kepler field of view, using the method of nonlinear time series analysis and higher order specral analysis. We conclude that a large majority of them show deterministic nonlinearity and low dimensional chaos as evidenced by the saturating correlation dimension and wide $f(\alpha)$ spectra. We look for the coupling between the eclipsing frequency and other frequencies in the system, using the main peak bicoherence function, $b_F(f)$, for the eclipsing frequency, $F$. We observe that the coupling is with the overtones of $F$, as well as with a number of other frequencies indicating nonlinear and chaotic dynamics in many binaries. To the best of our knowledge, this is the first study reported that suggests that nonlinear dynamics may be responsible for the irregular behavior of non compact binary systems.

We find correlations between nonlinear measures and astrophysical parameters, especially the fill out factor, $ff$. The fill-out factor is often associated with the evolution of the system, with a higher $ff$ implying a more evolved system. Hence this correlation seems to suggest that as a system evolves, the dynamics also undergo changes. Correlation dimension ($D_2$) is positively correlated with the fill-out factor, $ff$, while $SBF$ is negatively correlated with $ff$. Hence the fractal dimension of the system increases as the system evolves, whereas the coupling of other frequencies with the eclipsing frequency decreases. The correlations of $ff$ with $D_2$, $SBF$ and $\alpha_{min}$ are more pronounced in subcategories of binaries with specific ranges of mass ratio, $q$ and effective temperature, $T_{eff}$. This seems to suggest that these quantifiers can be used to predict the value for $ff$ in the restricted subcategories considered.

The additional constraints set by these correlations can help improve existing models of eclipsing binary stars. The increase in $D_2$ and decrease in $nmsd$ and $SBF$ with $ff$ indicates an increase in the stochasticity or turbulence in the system or changes in the parameters of the system. Similarly either noise or a change in the order of the dominant non-linearity could explain why the $SBF$ falls with $ff$. These changes can be attributed to relavent physical processes that become more prominent as the star evolves, like varying levels of spot activity or matter exchanges.

The establishment of low-dimensional chaos and its correlation to the intrinsic physical properties of the stars themselves seem to suggest a novel way to sub-classify these systems based on their nonlinear properties. Existing classifications rely mostly on the observational properties of the stars, with little regard for the nature of the underlying dynamics. Studying the effects of the various physical phenomena on nonlinear measures, can also help to narrow down on the causes behind unequal maxima in a particular binary star system. Further these correlations could be used in the prediction of the fill-out factor of these overcontact binary stars. The tools of machine learning are being used increasingly to predict the values for intrinsic parameters for large datasets like those from Kepler \cite{prvsa2011kepler,prvsa2008artificialint}. Nonlinear time series quantifiers could provide a complimentary approach to accurately predict the value of fill-out factor. Any model that attempts to re-create binary star light curves must be able to reproduce these correlations, which exist in the real systems. Hence these may serve to be a good check to determine the accuracy of binary star models. We believe that this work will be a leading edge into the productive application of the tools of nonlinear time series analysis into understanding the underlying physical processes governing contact binary stars.

\section{References}
\label{sec-7}

%\begin{thebibliography}{00}


%\bibitem{}

%\end{thebibliography}
\end{document}